\documentclass[preprint]{aastex}
\usepackage{pdflscape}
\usepackage{afterpage}
\usepackage{capt-of}
\usepackage{rotating}
\usepackage{threeparttable}
\usepackage{amsmath}
\usepackage{amssymb}
\usepackage{url}
\usepackage{graphicx}
\newcommand{\be}{\begin{equation}}
\newcommand{\ee}{\end{equation}}
\newcommand{\bea}{\begin{eqnarray}}
\newcommand{\eea}{\end{eqnarray}}

\usepackage{color}              
\usepackage{epsfig}
\usepackage{amsmath}
\usepackage{amssymb}
\usepackage{amstext}

\graphicspath{
{figures/}
}
\begin{document}

\title{First detection at 5.5 and 9 GHz of the radio relics in bullet cluster with ATCA}
\author{Siddharth Malu\altaffilmark{1}, Abhirup Datta\altaffilmark{1,2} and Pritpal Sandhu\altaffilmark{1}}
\altaffiltext{1}{Centre of Astronomy, Indian Institute of Technology Indore, Simrol, Khandwa Road, Indore 452020, India}
\altaffiltext{2}{Center for Astrophysics and Space Astronomy, Department of Astrophysical and Planetary Science, University of Colorado, Boulder, C0 80309, USA}
\begin{abstract}
We present here results from observations at 5.5 and 9 GHz of the Bullet cluster 1E 0657$-$55.8 with the
Australia Telescope Compact Array (ATCA). Our results show detection
of diffuse emission in the cluster. Our findings are consistent with
the previous observations by \citet{2014MNRAS.440.2901S} \&
\citet{2015MNRAS.449.1486S} at  1.1--3.1 GHz. Morphology of diffuse
structures (relic regions A and B and the radio halo) are consistent with those reported by the previous study. Our results indicate steepening in the spectral index at higher frequencies ($\gtrsim$ 5.0 GHz) for region A. The spectrum can be fit well by a broken power law. We discuss the possibility of a few recent theoretical models explaining this break in the power law spectrum, and find that a modified Diffusive Shock Acceleration (DSA) model or a turbulent
reacceleration model may be relevant. Deep radio observations at high frequencies ($\gtrsim$ 5 GHz) are required for a detailed comparison with this model.
\end{abstract}


\section{Introduction}
\label{intro}
Galaxy clusters have been studied extensively in X-rays \citep{1988xrec.book.....S} and radio \citep{2014IJMPD..2330007B,2012A&ARv..20...54F}. Galaxy clusters grow by mergers with other clusters and galaxy groups. These mergers create shock waves within the
ICM that can accelerate particles to extreme energies, with the compression of magnetic fields along shock fronts
causing an additional systematic pressure enhancement. These shocks can be studied through two different processes -- Brehmsstrahlung (through X-ray observations) and Synchrotron (through radio), and they probe thermal and non--thermal populations of electrons.

Diffuse non-thermal radio emission in clusters of galaxies of
$\sim$Mpc size not associated with galaxies, when close to the centres of
clusters, are called {\em radio halos}, and when at or close to the
peripheries of clusters are called {\em radio relics}. These are associated with 
relativistic electrons and magnetic fields in the ICM.

1E0657$-$56, known as the `Bullet Cluster', is one of the hottest
known clusters (with X-ray luminosity $L_{X}\sim$4.3$\times$10$^{45}$ ergs s$^{-1}$; temperature $kT\sim$ 14.7 keV) that has been well-studied over the last decade for a
variety of reasons; namely, the existence of a cold front in the X--ray observations \citep{2009ApJ...704.1349O} a strong radio halo \citep{2000ApJ...544..686L}, the Sunyaev-Zel'dovich effect (\citet{2009ApJ...701...42H}, \citet{2010ApJ...716.1118P} and
references therein), though most notably in providing the most direct proof of the existence of dark matter \citep{2006ApJ...648L.109C}. It is
a cluster collision/merger event at $z\sim$0.296, with the larger,
eastward cluster being $\sim$10 times the mass of the smaller `bullet'.  

A powerful radio halo in the Bullet cluster was first reported by \citet{2000ApJ...544..686L}, 
who detected the radio halo using
the ATCA as well as 843 MHz Molonglo Observatory Synthesis Telescope
(MOST). Analysis of data in \citet{2000ApJ...544..686L} centered on a 3.5 Mpc$^{2}$ region defined
by them on the basis of the extent of diffuse emission observed at 1.3 GHz.

\citet{2014MNRAS.440.2901S} reported deep observations of the cluster
at 2.1 GHz, with a noise rms of 15$\mu$Jy beam$^{-1}$, a significant
improvement over \citet{2000ApJ...544..686L}. This was made possible
by two upgrades of the Australia Telescope Compact Array (ATCA) -- addition of a N--S spur, and the increase in bandwidth by a factor of 16, through the Compact Array Broadband Backend, or CABB \citep{2011MNRAS.416..832W}.
These two upgrades improved the sensitivity of ATCA by a factor of
$\sim$4, and enabled deep observations of Bullet cluster at 2.1 GHz
and higher frequencies. 

\citet{2014MNRAS.440.2901S} constructed a spectral index map of the
Bullet cluster, and detected polarization in a certain
region. \citet{2015MNRAS.449.1486S} studied this region in more
detail, and confirmed that it has the characteristics of a relic. They
then proceeded to estimate the total brightness of the two components
marked `Region A' \& `Region B' in four sub--bands, and derive spectral
indices $-$1.07$\pm$0.03 and $-$1.66$\pm$0.14 for the two components
respectively. They also interpret `Region B' as a second shock front. One of the most
important aspects of the detection and characterization of diffuse
emission in \citet{2014MNRAS.440.2901S} and
\citet{2015MNRAS.449.1486S} is their ``in--band'' spectral index
estimation, i.e. determination of spectral index within the 1.1--3.1
GHz ATCA band. 

2.1 GHz deep observations of \citet{2014MNRAS.440.2901S} and
\citet{2015MNRAS.449.1486S} are critical
not only for an accurate estimation of the radio halo / relic flux,
but also to accurately determine the spectral index of both the radio
halo and the radio relic. As a practical matter, it is in fact easier
to estimate the total flux and spectral index of the radio relic as
compared to the radio halo, since halos, due to their central location
in cluster mergers, are likely to be in the vicinity of several point
sources (bright radio galaxies). 

We summarize our observations in \S\ref{radio_obs}, describe data reduction in \S\ref{datareduction}, present results in \S\ref{results} (5.5 GHz results in \S\ref{radioimage1}, 9 GHz results in \S\ref{radioimage2} and the radio relics in \S\ref{relic}), and discuss the results in \S\ref{discussion}.

\begin{figure}[!h]
	\includegraphics[width=\columnwidth,angle=0]{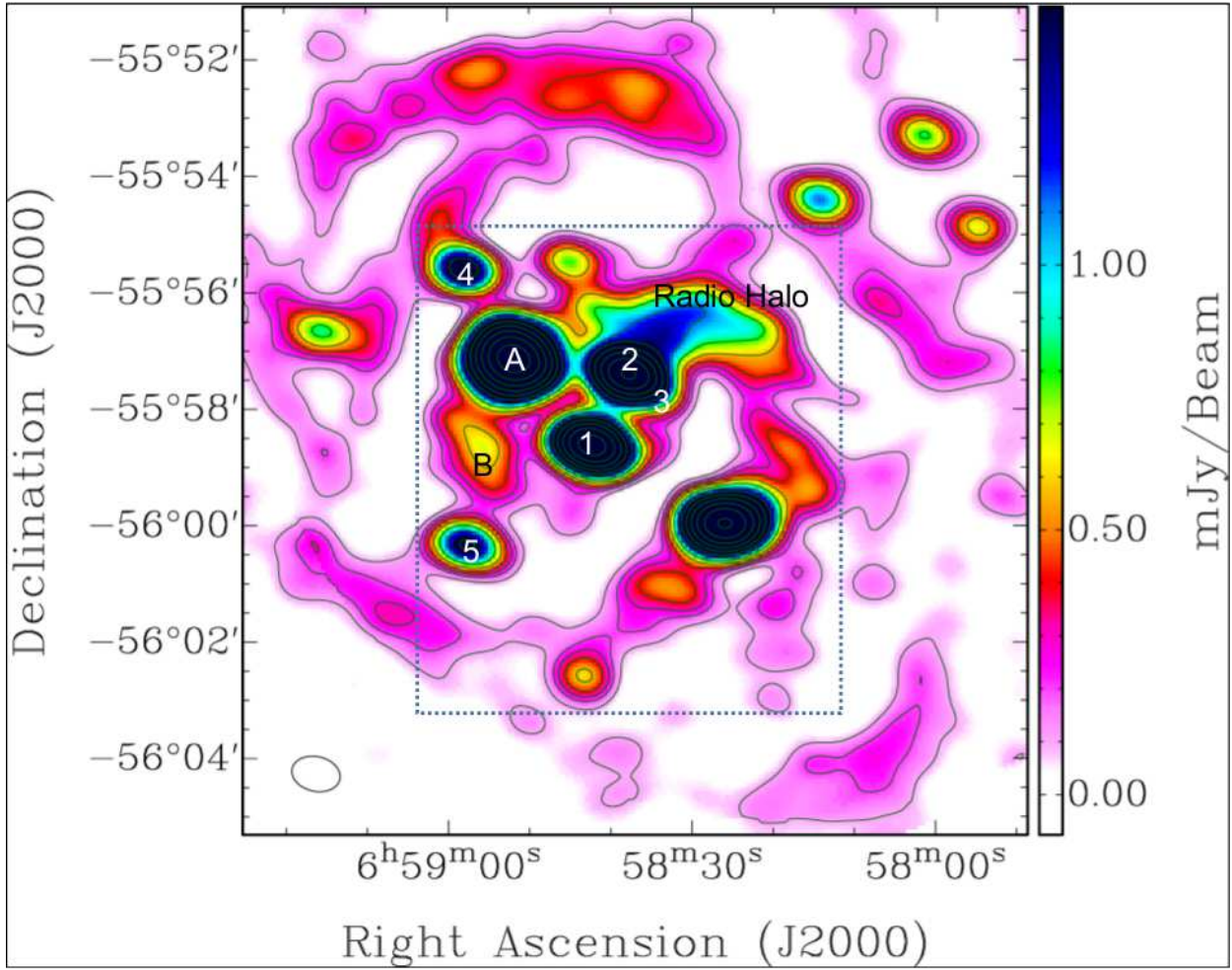}
    \caption{5.5 GHz radio image of the Bullet cluster,
      observed with H168 array of the ATCA, with a
      synthesized beam 50.5$\arcsec\times$35.9$\arcsec$ (indicated as an
      ellipse in the bottom left corner). This is a natural--weighted image, and the extent of this
      image is similar to the figures in \citet{2014MNRAS.440.2901S}. Noise rms ($\sigma_{\mathrm{RMS}}$) is 20$\mu$Jy/beam. Contour levels are at $-5,5,10,15,20,40,80,160,320,640\mathrm{~}\times\mathrm{~}\sigma$.  }
    \label{5ghzfig01}
\end{figure}

\begin{figure}[!h]
	\includegraphics[width=\columnwidth,angle=0]{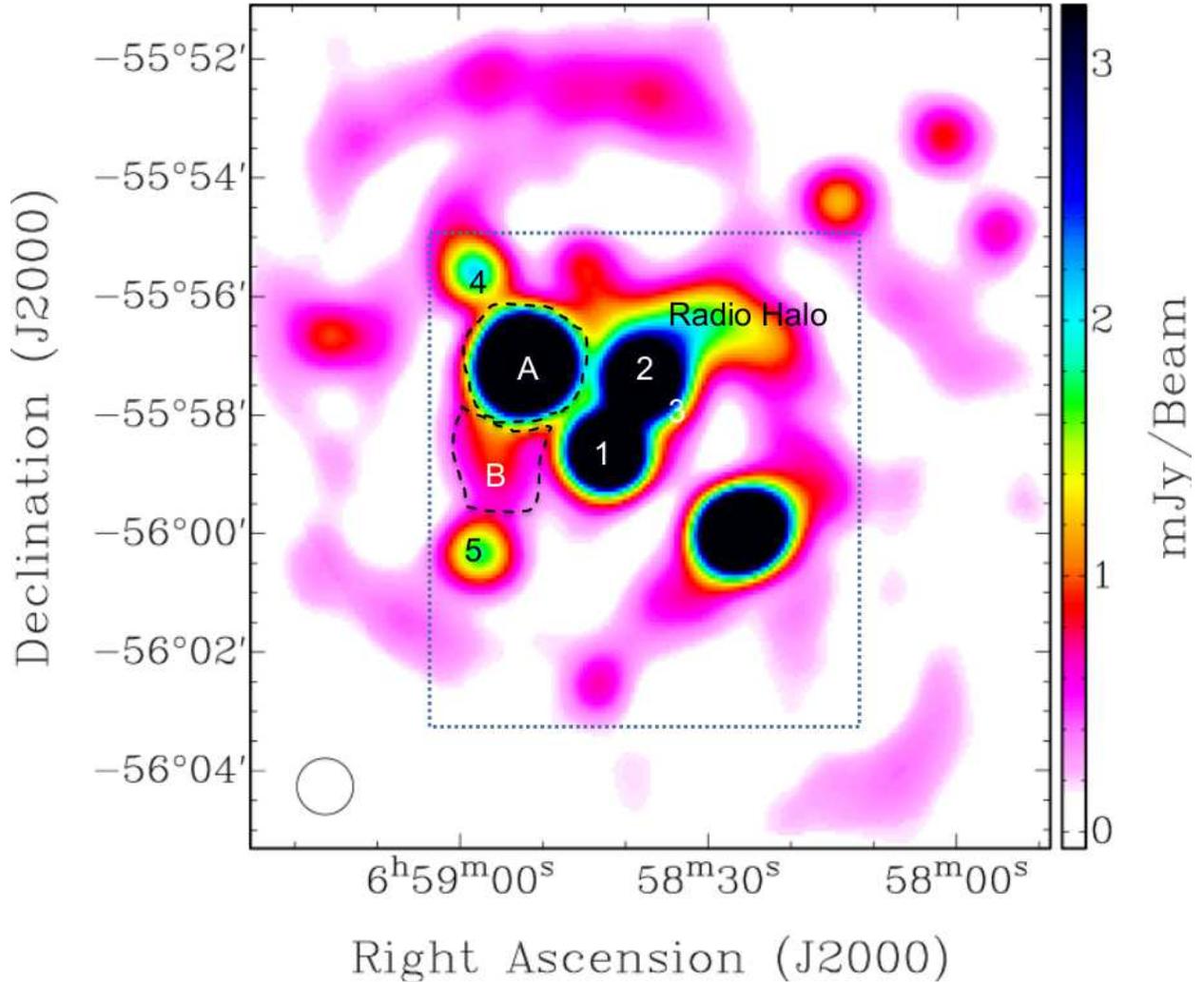}
    \caption{5.5 GHz radio image of the Bullet cluster,
      observed with H168 array of the ATCA, with a
      synthesized beam 57$\arcsec\times$57$\arcsec$ (indicated as an
      ellipse in the bottom left corner). This is a natural--weighted image, and the extent of this
      image is similar to the figures in
      \citet{2014MNRAS.440.2901S}. Noise rms ($\sigma_{\mathrm{RMS}}$)
      is 20$\mu$Jy/beam. Regions A \& B, and the six point sources are
      marked in the figure. Properties of the point sources and
      regions A \& B are given in Tables ~\ref{ptsrc} and
      \ref{spectrum1}. The dashed box indicates the area in the image
      that we have described in this paper. The two dashed areas indicate roughly the extents of the regions A and B.}
    \label{5ghzfig02}
\end{figure}

\section{Observations}
\label{radio_obs}
The Australia Telescope Compact Array (ATCA)  is a radio
interferometer with six 22m antennas, five of which may be positioned
on stations along a T-shaped rail track that is 3-km along E-W and
214-m along N-S.

The bullet cluster was observed for a total duration of 14 hours in
the 6 cm and 3 cm bands with centre frequencies 5.5 GHz and 9 GHz
respectively, in a 2--pointing mosaic, with the pointing
centers at (J2000 epoch coordinates) RA: $06^{\rm h}58^{\rm m}30^{\rm s}$, 
DEC: $-55\degr57\arcmin00\arcsec$ and RA: $06^{\rm h}58^{\rm m}20^{\rm s}$, 
DEC: $-55\degr56\arcmin00\arcsec$ respectively. Table~\ref{obs_journal} lists the total amount of time spent on these two pointings, and in slewing and calibration. Imaging from visibilities in this mosaic is described in \S\ref{datareduction}. 

\begin{center}
\begin{table}
\caption{Summary of the ATCA observations}
\label{obs_journal}
\begin{tabular}{@{}llcr}
\hline
Array & Frequency & Observing    & Date \\
      & (GHz)     & time (hours) & \\
\hline
H168 & 5.5 & 10.0 & 2010 July 30 \\
H168 & 5.5 & 4.0  & 2010 July 31 \\
H168 & 9.0 & 10.0 & 2010 July 30 \\
H168 & 9.0 & 4.0  & 2010 July 31 \\
\hline
\end{tabular}
\end{table}
\end{center}

Observations were made in the H168 array that has a maximum baseline up to
192-m (using the five antennas in the array) and a minimum baseline of
61-m. A summary of the observations is in Table 1. Observations of
the Bullet cluster were made in a pair of 2-GHz bands: a `5.5-GHz band'
covering frequencies 4.5--6.5 GHz and a `9-GHz band' covering the range
8--10 GHz. Each of the 2-GHz wide bands were subdivided into 2048 frequency
channels. All observations were in full polarization mode and recorded
multi-channel continuum visibilities. In each observing session,
antenna pointing corrections were updated every hour using a 5--point
offset pattern observation on a bright calibrator, unresolved phase
calibrators were observed every 60 min to monitor and correct for
amplitude and phase drifts in the interferometer arms, and PKS
B1934--638 was observed once every session as a primary calibrator to
set the absolute flux density scale. Visibilities were recorded with
10 sec averaging. 

PKS~B1934$-$638 was used as the primary calibrator, and PKS~B0823$-$500 was used as a phase \& bandpass calibrator.

The synthesized beam is much larger than in \citet{2014MNRAS.440.2901S} because of our observations being made in the
second--most compact configuration at the ATCA, namely, H168. This restricts the longest baseline to 168-m, or $\sim$3k$\lambda$.

The Compact Array Broadband Backend, described in
\citet{2011MNRAS.416..832W}, was used, with two independent 2048 MHz windows (dual
polarization) for correlation, with 1 MHz resolution, which is the standard setting for continuum observations.

\section{Data Reduction}
\label{datareduction}
Data were analysed using the Multichannel Image Reconstruction, Image
Analysis and Display (MIRIAD, developed by ATNF-CSIRO); all
image processing were also accomplished using utilities in this software package.

Adopted fluxes for the primary calibrator PKS~B1934$-$638 were 4.965
and 2.700 Jy in the 6-cm and 3-cm bands respectively; the spectral index was adopted to be $-1.23$ in both
bands (see \citet{atca_sault2003}). Outliers in the amplitudes of visibility data on PKS~B1934$-$638 were
rejected---removing $\sim$10\% of data---and the reliable visibilities
were used to set the absolute flux density
scale as well as determine the instrument bandpass calibration.

When calibrated for the bandpass, the visibility amplitudes of PKS~B0823$-$500
showed continuity across both the 2-GHz observing frequency ranges and a trend
consistent with a single power-law: this was a check of the bandpass
calibration.  Drifts of up to $30\degr$ were observed in the interferometer arms
over the observing sessions: calibrations for the time-varying complex
gains in the antenna signal paths as well as calibrations for
polarization leakages were derived from the visibilities on
PKS~B0823$-$500. RMS phase variations in antenna signal paths within
the 1-min calibrator scans was within $0.5\degr$, indicating that
short timescale atmospheric and instrumental phase cycling would result in amplitude attenuation of
less than 1\%. 

The visibility data in the 5.5 and 9 GHz bands were separately edited for
interference and calibrated before bandwidth synthesis imaging.
Visibility data in each of the 2 GHz wide bands were recorded over 2048 frequency channels, and 50 channels at
each of the band edges were excluded from analysis to avoid data in frequency
domains where signal path gains are relatively low. Frequency channels that
appeared to have relatively large fluctuations in visibility amplitude owing to hardware 
faults in the digital correlator were also rejected prior to calibration and imaging.

Since no circular polarization is expected, Stokes V is expected to be consistent with
thermal noise, and we make estimates of noise rms from Stokes V, following \citet{2000MNRAS.315..808S}. Therefore, at times and frequency channels where Stokes V
visibilities deviate more than four times rms thermal noise in the calibrated
visibilities acquired towards the cluster pointings, data in all
Stokes parameters were rejected. Stokes--V based clipping was
therefore done, aimed at automated rejection of self-generated low
level interference. In order to carry out Stokes--V based clipping,
the {\sc Miriad} task {\sc Tvclip} was used to find the median and rms
levels for every channel, and then all visibilities beyond the median
$\pm$5$\sigma$ were flagged, and the same flagging was applied to all
the other Stokes parameters. This procedure was then repeated for every Stokes
parameter. This way, 25\% of the data at 5.5 GHz and 20\% of the data
at 9 GHz were flagged. 

We note here that our observations were made in 2010, when the CABB
system \citep{2011MNRAS.416..832W} was relatively new, and therefore, noise per channel
was higher, resulting in a higher T$_{\rm{sys}}$, i.e. system temperature/lower sensitivity
than is currently possible. Going through our raw ATCA data, we find
that the typical T$_{\rm{sys}}$ in our data is $\sim$ 80--100K, whereas
current estimates of T$_{\rm{sys}}$ are in the range 40--50
K.

\begin{center}
\begin{sidewaystable}
\begin{threeparttable}
\caption{Unresolved and diffuse continuum radio sources detected in the Bullet cluster field}
\label{ptsrc}
\begin{tabular}{lcccccclrrc}
\hline\hline
Source & \multicolumn{6}{c}{RA~~~~~~~~~~~~DEC} & Diffuse or
& {Int. Flux Density} & {Int. Flux Density} & \citet{2014MNRAS.440.2901S}\\
Label & \multicolumn{6}{c}{(J2000)} & point source & $S_{\rm 5.5~GHz}$ & $S_{\rm 9~GHz}$ & Label\\ \hline  & h & m & s & $^\circ$ & $\arcmin$ & $\arcsec$ &  &  (mJy) &  (mJy) & \\ 
\hline
Region A & 06 & 58 & 51.91 & $-$55 & 57 & 07.9 & Diffuse & 18.04 $\pm$ 0.40 & 4.36$\pm$0.15 & Region A\\
Region B & 06 & 58 & 55.26 & $-$55 & 58 & 47.8 & Diffuse &  0.88 $\pm$ 0.08 & 0.09$\pm$0.02 & Region B \\
1 & 06 & 58 & 42.3 & $-$55 & 58 & 37.5 & Point & 8.69 $\pm$ 0.20 & 2.65 $\pm$ 0.11 & M\\
2 & 06 & 58 & 37.6 & $-$55 & 57 & 24.0 & Point & 7.54 $\pm$ 0.18 & 4.36 $\pm$ 0.13 & L\\
3 & 06 & 58 & 34.3 & $-$55 & 57 & 40.0 & Point & 0.60 $\pm$ 0.10 & Detected & K\\
4 & 06 & 58 & 58.1 & $-$55 & 55 & 35.1 & Point & 2.16 $\pm$ 0.02 & 0.43 $\pm$ 0.02 & Detected\\
5 & 06 & 58 & 57.8 & $-$56 & 00 & 20.1 & Point & 1.89 $\pm$ 0.02 & 0.39 $\pm$ 0.04 & Detected\\ 
6 & 06 & 58 & 14.6 & $-$55 & 54 & 23.0 & Point & 0.99 $\pm$ 0.01 & 0.75$\pm$ 0.04 & N\\ \hline\hline
\end{tabular}
\begin{tablenotes}[flushleft]
\item {\sc notes}-- Letters indicate diffuse sources; numbers indicate
  discrete (point) sources from our 5.5 \& 9 GHz observations. Regions
  of diffuse emission are marked with letters on
  Fig.~\ref{5ghzfig02} and Fig.~\ref{9ghzfig02}. 
\end{tablenotes}
\end{threeparttable}
\end{sidewaystable}
\end{center}

In order to mosaic the two pointings, we follow the scheme provided by
\citet{1996A&AS..120..375S}; namely, make a mosaic of all the data,
leading to a mosaicked dirty image, using the option `mosaic' in the
{\sc Miriad} task {\sc Invert}. Then, deconvolution was done using
{\sc Miriad} task {\sc Mossdi}, which allows a joint deconvolution of mosaicked images.

\begin{figure}
	\includegraphics[width=\columnwidth,angle=0]{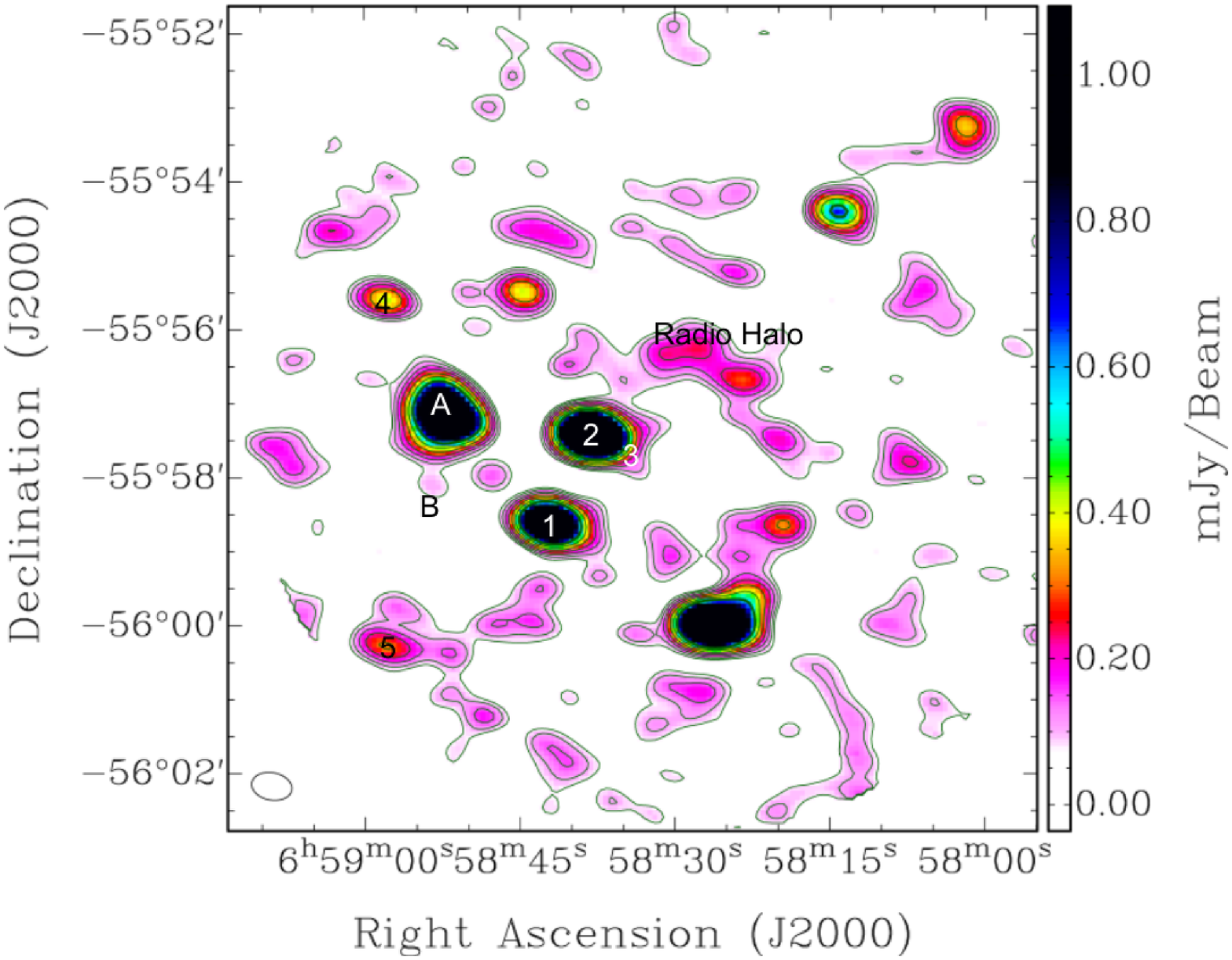}
    \caption{9 GHz radio image of the Bullet cluster,
      observed with H168 array of the ATCA, with a
      synthesized beam 33.1$\arcsec\times$22.3$\arcsec$ (indicated as an
      ellipse in the bottom left corner). This is a natural--weighted image, and the extent of this
      image is similar to the figures in
      \citet{2014MNRAS.440.2901S}. Noise rms ($\sigma_{\mathrm{RMS}}$)
      is 15$\mu$Jy/beam. Contour levels are at $-5,3,5,7.1,10,14,20,28,40\mathrm{~}\times\mathrm{~}\sigma$. Regions A \& B, and the
six point sources are marked in the figure.}
    \label{9ghzfig01}
\end{figure}

\begin{figure}
	\includegraphics[width=\columnwidth,angle=0]{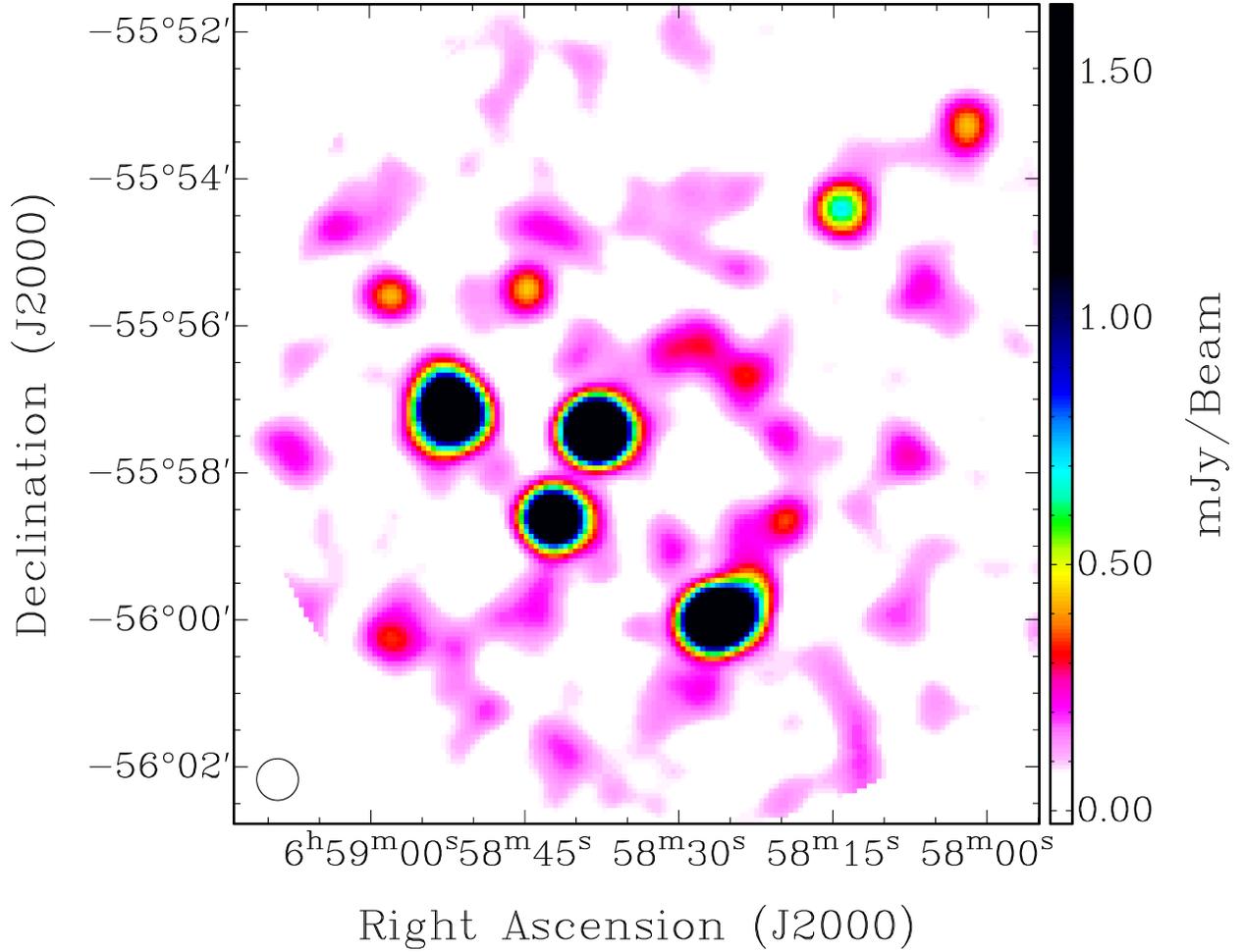}
    \caption{9 GHz radio image of the Bullet cluster,
      observed with H168 array of the ATCA, with a
      synthesized beam 34$\arcsec\times$34$\arcsec$ (indicated as an
      ellipse in the bottom left corner). This is a uniform--weighted image, and the extent of this
      image is similar to the figures in \citet{2014MNRAS.440.2901S}. Noise rms ($\sigma_{\mathrm{RMS}}$) is 25$\mu$Jy/beam.}
    \label{9ghzfig02}
\end{figure}

\section{Results $-$ continuum imaging}
\label{results}
\subsection{Radio Image at 5.5 GHz}
\label{radioimage1}
Following the analysis described in previous sections, we obtained a
radio image of the bullet cluster field, which is shown in
Fig.~\ref{5ghzfig01}. This image was made with natural weighting, and
so were all the subsequent images. It should be noted that the
intrinsic resolution of this image is
50.5$\arcsec\times$35.9$\arcsec$. A list of point sources, and of
regions with diffuse emission, with their location and total emission, is given in Table~\ref{ptsrc}; locations of all point sources match with the values in \citet{2014MNRAS.440.2901S}. 

There is clear evidence of diffuse emission at 5.5 GHz, which occurs in the same region as the 2.1 GHz image (Fig. 5) in
\citet{2014MNRAS.440.2901S}. To facilitate comparison with the 2.1 GHz
observations in \citet{2014MNRAS.440.2901S}, another figure has been made with a synthesized beam of 57$\arcsec\times$57$\arcsec$. Most
baselines in the \citet{2014MNRAS.440.2901S} 2.1 GHz image are
$\gg$5k$\lambda$, so it is not surprising that the diffuse
emission we find in the 5.5 GHz image is similar as compared
to 2.1 GHz \citep{2014MNRAS.440.2901S} -- morphologically. The
total amount of diffuse emission in these regions is stated in
Table~\ref{ptsrc}.

There is clear detection of the radio halo and two components of the radio relic, as described in \citet{2014MNRAS.440.2901S} and \citet{2015MNRAS.449.1486S} -- these are clearly marked in Fig.~\ref{5ghzfig02}. 

Qualitatively, diffuse emission in Fig.~\ref{5ghzfig02} is in good
agreement with the 2.1 GHz image (Fig. 5) in \citet{2014MNRAS.440.2901S} --
both in terms of extent and morphology. In particular, the radio halo
at 5.5 GHz has a similar morphology and extent as the halo at 2.1 GHz in \citet{2014MNRAS.440.2901S}. We also
detect -- at relatively low significance -- a bridge between the radio
halo and Region A. Like \citet{2014MNRAS.440.2901S}, we
detect more than one local maxima in the radio halo, and two of these
local maxima coincide with the X--ray brightness centers, as in
the 2.1 GHz image. The radio halo is extended along the merger axis,
the physical extent of the radio halo is more along the merger axis
than perpendicular to it, and there is a well--defined western edge, again, like
the 2.1 GHz image in \citet{2014MNRAS.440.2901S}. 

\subsection{Radio Image at 9 GHz}
\label{radioimage2}
A low--resolution natural--weighted image at 9 GHz of the Bullet cluster, with a
synthesized beam of 33.1$\arcsec\times$22.3$\arcsec$ is shown in
Fig.~\ref{9ghzfig01}, and clearly shows diffuse emission is several
regions -- region A, region B and radio halo. Another image, with synthesized beam
34$\arcsec\times$34$\arcsec$ (Fig.~\ref{9ghzfig02}) was made to
compute the spectral index of regions A \& B. Region A is very
prominent, but region B is barely detectable, and part of region B are
absent, implying that the spectrum might have steepened.

There are several similarities with the 2.1 and 5.5 GHz images; for
instance, just as in \citet{2014MNRAS.440.2901S} and \S\ref{radioimage1}, we
detect more than one local maxima in the radio halo, and two of these
local maxima coincide with the X--ray brightness centers, just as in
the 2.1 and 5.5 GHz images -- the third local maxima coinciding with
the local maxima in radio halo at 5.5 GHz. The radio halo is extended along the merger axis,
just as at 2.1 and 5.5 GHz, and there is a well--defined western edge, just as
in the 2.1 and 5.5 GHz images. In other words, features of the radio halo at 9 GHz match qualitatively with the features detected by \citet{2014MNRAS.440.2901S}.

\subsection{Radio relic at 5.5 and 9 GHz}
\label{relic}
\citet{2015MNRAS.449.1486S} have detected and characterized relics in the Bullet
cluster at 2.1 GHz. They measured the brightness, substructure and
polarization properties of the two relic regions. They
interpret Region B as a second shock front, and also comment on the high
brightness of Region A, and speculate on its origin.  

In this paper, we have detected both relic regions A and B in our 5.5 and 9 GHz observations (Figs.~\ref{5ghzfig01}, \ref{5ghzfig02}, \ref{9ghzfig01} \& \ref{9ghzfig02}). Peak emission occurs at  RA: $06^{\rm h}58^{\rm m}51.91^{\rm s}$, DEC:
$-55\degr57\arcmin7.88\arcsec$ at 5.5 GHz, and at RA: $06^{\rm h}58^{\rm m}51.91^{\rm s}$, DEC:
$-55\degr57\arcmin11.88\arcsec$ at 9 GHz, the difference between the
two positions being 4$\arcsec$, which is a fraction of the
beamsize. Our definitions of Relic Regions `A' and `B' are exactly the same as in
\citet{2015MNRAS.449.1486S} -- with the same extents in declination. 

\begin{table}
\caption{Diffuse emission in Relic Regions A \& B in C and X bands}
\label{spectrum1}
\begin{tabular}{@{}lcc}
\hline
Frequency & Region A & Region B \\
 (MHz)    & (mJy)    & (mJy) \\
\hline
5021 & 19.05 $\pm$ 0.20 & 0.88 $\pm$ 0.08 \\
5468 & 18.04 $\pm$ 0.20 & 0.78 $\pm$ 0.07 \\
5944 & 12.86 $\pm$ 0.13 & 0.66 $\pm$ 0.05 \\
9013 & 04.36 $\pm$ 0.15 & 0.09 $\pm$ 0.02 \\
\hline
\end{tabular}
\end{table}

\begin{figure}
  \includegraphics[width=\columnwidth,angle=0]{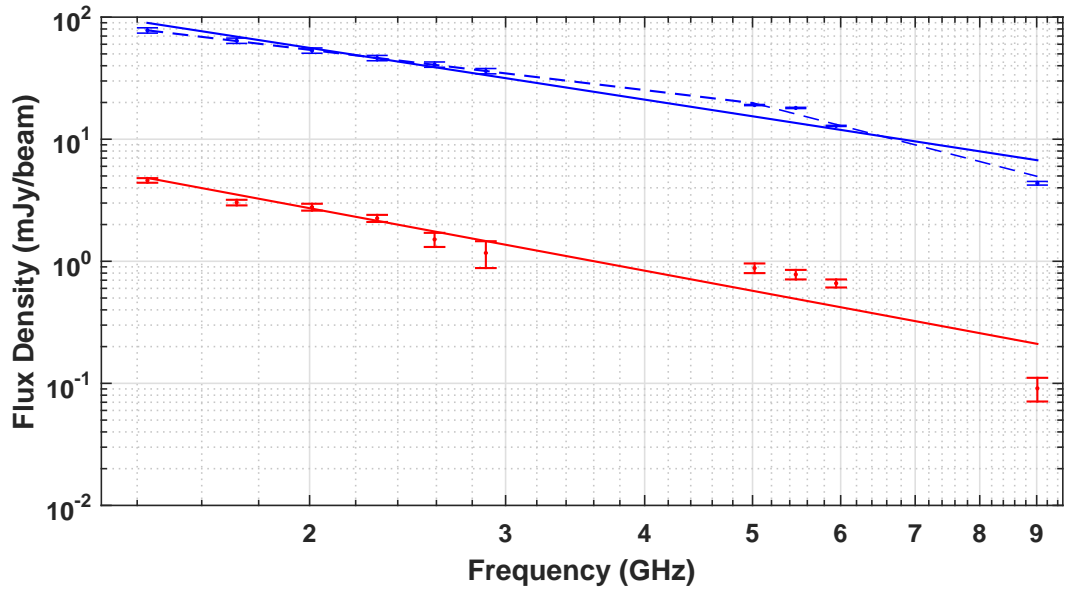}
  \caption{Spectral Energy Density of Relic Regions A and B between 1.1 and 9 GHz. Data points for 5.0 and 5.9 GHz are taken from Fig.~\ref{5ghzfig02} and data points for 9.0 GHz are taken from Fig.~\ref{9ghzfig02}. Data points between 1.1 and 3.1 GHz are taken from Fig. 2 of \citet{2015MNRAS.449.1486S}. Blue represents Region A and red represents Region B. The solid blue line indicates a single spectral index fit for region A, and dashed blue lines represent a 2--index model for region A. See the discussion in \S\ref{relic}.}
  \label{cxspectrum1}
\end{figure}

The general large--scale morphology of Regions A and B is roughly
similar to figures 1 \& 3 in \citet{2015MNRAS.449.1486S}; however, our
observations lack angular resolution, so it is not possible to tell
whether the high--resolution structure at 5.5 and 9 GHz is similar to
the 2.1 GHz structure.

We have measured the integrated flux of the diffuse radio emission as a
function of frequency for regions A and B (see Fig.~\ref{cxspectrum1})
and characterized the 4.5 to 6.5 GHz \& 8 to 10 GHz emission. Following \citet{2015MNRAS.449.1486S}, we calculated the
uncertainty on our integrated flux density measurements by adding in
quadrature the ATCA absolute flux calibration error of 2 per cent
\citep{reynolds1994}, with the error on the integrated flux density
derived from the image noise. Flux densities for regions A and B at
5.021, 5.468, 5.944 and 9.013 GHz are provided in Table~\ref{spectrum1}.

In computing the spectral index, we have taken also taken into account
the flux densities of regions A and B as presented in Fig. 2 in
\citet{2015MNRAS.449.1486S}. Following Fig.~\ref{cxspectrum1}, we
obtain the mean spectral index of region A to be
$-$1.41$^{+0.09}_{-0.07}$ between 1.1 and 9.0 GHz. For region B, we obtain a mean spectral
index of $-$1.70$^{+0.35}_{-0.31}$ between 1.1 and 9.0 GHz. For region A, we have also fit for
a 2--spectral index model: one between 1.1 and 5.0 GHz and the other
between 5.0 and 9.0 GHz. 1.1 to 3.1 GHz data is from \citet{2014MNRAS.440.2901S} and \citet{2015MNRAS.449.1486S}. The two mean spectral indices are
$\alpha^{1.1}_{5.0}=-$1.1 and $\alpha^{5.0}_{9.0}=-$2.4. We present
the spectrum of Relic Regions A and B in the 1.1--10 GHz range, in Fig.~\ref{cxspectrum1}. 

As can be seen from Fig.~\ref{cxspectrum1}, the 2-spectral index model fits the data really well, which hints at the existence of a break in the power law between 1 and 9 GHz. A similar steepening was reported in \citet{2014MNRAS.441L..41S} for the `Sausage' cluster. It should be noted that such high--frequency steepening is inconsistent with Diffusive Shock Acceleration theory \citep{1983RPPh...46..973D}. \citet{2014MNRAS.441L..41S} state that high-frequency steepening could be caused by an inhomogeneous medium with temperature/density gradients, or by lower acceleration efficiencies of high-energy electrons. This is discussed in \S\ref{discussion}. 

\subsection {A note on the radio halo at 5.5 and 9 GHz}
\label{radiohalo}
As mentioned in \S\ref{radioimage1} and \S\ref{radioimage2}, the radio
halo is easily detectable at both 5.5 and 9 GHz in our observations,
and we also detect several peaks in the radio halo, spread along the
merger axis of the cluster. Despite this, we have not done a detailed analysis at 5.5 and 9 GHz, because our observations lack the
angular resolution to identify and characterize point sources, in order to
exclude them from estimates of diffuse emission flux. The only details we
provide here are as follows. Integrated flux of the radio halo within the
5$\sigma$ contours, which is 7.1$\pm$0.3 mJy at 5.5 GHz and
1.6$\pm$0.3 mJy at 9 GHz. 
We do not compute the spectral index for the radio halo, since
point--sources were not subtracted to get the above
estimates. Improved data at 5.5 and 9 GHz with other ATCA
configurations are essential for estimating the spectral index of the
radio halo. We have applied for more time on the ATCA, with arrays that will provide higher resolution.

\subsection {Point sources at 5.5 and 9 GHz}
\label{pointsources1}
We present detection of 6 point sources in the Bullet cluster
field at 5.5 and 9 GHz. Each one of these
displays interesting features, especially the possible presence of
jets. Sources 1,2,3,6, whose properties are described in Table 2 in
\citet{2014MNRAS.440.2901S}, were also detected by us; their
properties are listed in Table~\ref{ptsrc}.
%
%
\section{Discussion}
\label{discussion}
In this paper, we have presented the first detections of diffuse emission in the
Bullet cluster at 5.5 and 9 GHz, using the ATCA (radio halo and relic regions A \& B). We summarize the essential results as below:
\begin{enumerate}
\item\label{1} The brightest part of diffuse emission in the cluster is a
  northern relic (referred to in \citet{2015MNRAS.449.1486S}
  and in this paper as Region A) -- significantly brighter than all other
  diffuse features, and unusually bright at 5.5 and 9 GHz, like the 2.1 GHz observations
\item\label{2} There is a second shock front -- named Region B -- immediately
  to the south of Region A, which is elongated along N--S
\item\label{3} At very low resolutions (57$\arcsec$ at 5.5 GHz and
  34$\arcsec$ at 9.0 GHz), morphology of
  regions `A' and `B' are similar at 5.5 and 9 GHz compared to 2.1 GHz
\item\label{4} Morphology and extent of the radio halo in the cluster are
  similar at 5.5, 9.0 and 2.1 GHz
\item\label{5} Spectral index calculation shows a steepening at higher frequencies for Region A, which can best be characterized by a broken power law near 5.0 GHz. 
\end{enumerate}

Diffuse emission in cluster mergers has been detected at high frequencies ($>$ 5 GHz) for the Bullet
and other clusters \citep{2010arXiv1005.1394M,2011JApA...32..541M}; in
particular, \citet{2016MNRAS.455.2402S} have charaterized the spectrum
of the `Sausage' and `Toothbrush' radio relics \citep{2010Sci...330..347V,2012A&A...546A.124V} from 150 MHz to 30
GHz. Therefore, a detection of the radio relic in the Bullet cluster
at 5.5 and 9 GHz is unsurprising. 

While an ideal Diffusive Shock Acceleration (DSA) model (e.g. \citet{1983RPPh...46..973D}) does not predict a break in the spectrum, more realistic models
predict a gradual steepening in the 0.1--10 GHz range \citep{2015JKAS...48....9K,2016arXiv160307444K}. Breaks in the spectrum are usually observed to be in the 1--2 GHz. In \citet{2016MNRAS.455.2402S} especially, the breaks in the spectra of the `Sausage' and `Toothbrush' radio relics are in the range 2--2.5 GHz. In the case of the observations presented in this paper, the break occurs close to 5 GHz. It is worth pointing out that the possibility of the break in the spectrum of the Bullet cluster occuring in the range 4.8--8.8 GHz has been pointed out \citep{2004AcPPB..35.2131S}, based on DSA modeling of the radio relic. 

The change/steepening in the spectral index in DSA models with radiative cooling is $\sim$ 0.5 (e.g. \citet{1998A&A...332..395E}), whereas the change we measure for our observations are $\geq$ 1. In \citet{2016MNRAS.455.2402S}, the change in the spectral index is $\sim$ 0.5 for the `Toothbrush' relic, and $\sim$ 0.8 for the `Sausage' relic.
Therefore, it is safe to say that there may be variations in the spectral index change due to steepening. 

\citet{2015ApJ...815..116F} point out that the turbulent reacceleration of cosmic ray (CR) electrons behind the shock can explain the spectral steepening -- which cannot be explained by the standard DSA model -- in a few cluster mergers, including the Bullet cluster. Interestingly, they point out possible obscurations of relics by turbulent reacceleration-formed radio halos; while no obscuring is observed in our observations or \citet{2014MNRAS.440.2901S}, both these observations detect a diffuse emission `bridge' between the relic and the halo, which may be due to the reacceleration of the CR electrons behind the shock. 

In order to explain the structure of radio relic spectra, \citet{0004-637X-809-2-186} explored DSA models where a spherical shock impinges on a population of relativistic electrons, as shown in their Fig. 1, which details a shock front meeting an elongated structure of fossil relativistic electrons. On comparison with `Sausage' cluster data from \citet{2016MNRAS.455.2402S}, they were able to explain the break in the spectrum, but not the complete shape of the spectrum. Interestingly, \citet{2015MNRAS.449.1486S} have pointed out the possibility of fossil relativistic electrons leftover from a radio galaxy causing the bright relic A, so that the modeling done by \citet{0004-637X-809-2-186} is directly applicable.

In this context, \citet{0004-637X-823-1-13} have considered the model of radio relic formation that is perhaps most relevant for the radio relic in the Bullet cluster, wherein the merger shock goes through a small-size collection of fossil electrons -- this causes the spectrum to be steeper than it would be only through radiative aging. Using their simulations, \citet{0004-637X-823-1-13} were abe to reproduce the spectral curvature in the Sausage cluster as measured in \citet{2016MNRAS.455.2402S}. Figs. 5 \& 6 in \citet{0004-637X-823-1-13} show their models, with data points from \citet{2016MNRAS.455.2402S}, in good agreement with their model.
 
 In order to compare the spectrum of the radio relic with the models detailed by \citet{0004-637X-823-1-13} and \citet{2015ApJ...815..116F}, high-resolution maps of the Bullet cluster at 5.5 and 9 GHz are needed -- hence the need for deep observations at these two frequencies, using the appropriate arrays, which we are going to apply for. Also, deeper radio observations along with X-ray data near the relic regions will allow us to study radio and X-ray properties of shocks in order to understand the dynamics of this merger \citep{2014ApJ...793...80D}. 

In summary, our 5.5 \& 9 GHz data exhibit a steepening of the spectrum of the radio relic in the Bullet cluster. Recent theoretical models may lead to a better understanding of the origin of the radio relic in the Bullet cluster, as has been demonstarted for the `Sausage' relic -- further observations with high resolution are needed at 5.5 and 9 GHz to facilitate comparison with these models.

\section*{Acknowledgements}
The Australia Telescope Compact Array is part of the Australia
Telescope which is funded by the Commonwealth of Australia for
operation as a National Facility managed by CSIRO. SM is greatful to
Maxim Voronkov, James Urquhart and the research staff at ATCA/ATNF for
guidance and help with understanding the CABB system. We thank Mark
Wieringa, Ravi Subrahmanyan and D. Narasimha for useful
discussions. Analysis was made possible by a generous grant for Astronomy by IIT Indore. We thank the anonymous referee for their comments.

\bibliographystyle{natbib}
\bibliography{bullet4}

\begin{thebibliography}{}

\bibitem[{Brunetti} and {Jones}(2014){Brunetti} and
  {Jones}]{2014IJMPD..2330007B}
{Brunetti}, G. and {Jones}, T.~W. (2014).
\newblock {Cosmic Rays in Galaxy Clusters and Their Nonthermal Emission}.
\newblock {\em International Journal of Modern Physics D\/}, {\bf 23}, 30007.

\bibitem[{Clowe} {\em et~al.}(2006){Clowe}, {Brada{\v c}}, {Gonzalez},
  {Markevitch}, {Randall}, {Jones}, and {Zaritsky}]{2006ApJ...648L.109C}
{Clowe}, D., {Brada{\v c}}, M., {Gonzalez}, A.~H., {Markevitch}, M., {Randall},
  S.~W., {Jones}, C., and {Zaritsky}, D. (2006).
\newblock {A Direct Empirical Proof of the Existence of Dark Matter}.
\newblock {\em \apjl\/}, {\bf 648}, L109--L113.

\bibitem[{Datta} {\em et~al.}(2014){Datta}, {Schenck}, {Burns}, {Skillman}, and
  {Hallman}]{2014ApJ...793...80D}
{Datta}, A., {Schenck}, D.~E., {Burns}, J.~O., {Skillman}, S.~W., and
  {Hallman}, E.~J. (2014).
\newblock {How Much can we Learn from a Merging Cold Front Cluster? Insights
  from X-Ray Temperature and Radio Maps of A3667}.
\newblock {\em \apj\/}, {\bf 793}, 80.

\bibitem[{Drury}(1983){Drury}]{1983RPPh...46..973D}
{Drury}, L.~O. (1983).
\newblock {An introduction to the theory of diffusive shock acceleration of
  energetic particles in tenuous plasmas}.
\newblock {\em Reports on Progress in Physics\/}, {\bf 46}, 973--1027.

\bibitem[{Ensslin} {\em et~al.}(1998){Ensslin}, {Biermann}, {Klein}, and
  {Kohle}]{1998A&A...332..395E}
{Ensslin}, T.~A., {Biermann}, P.~L., {Klein}, U., and {Kohle}, S. (1998).
\newblock {Cluster radio relics as a tracer of shock waves of the large-scale
  structure formation}.
\newblock {\em \aap\/}, {\bf 332}, 395--409.

\bibitem[{Feretti} {\em et~al.}(2012){Feretti}, {Giovannini}, {Govoni}, and
  {Murgia}]{2012A&ARv..20...54F}
{Feretti}, L., {Giovannini}, G., {Govoni}, F., and {Murgia}, M. (2012).
\newblock {Clusters of galaxies: observational properties of the diffuse radio
  emission}.
\newblock {\em \aapr\/}, {\bf 20}, 54.

\bibitem[{Fujita} {\em et~al.}(2015){Fujita}, {Takizawa}, {Yamazaki},
  {Akamatsu}, and {Ohno}]{2015ApJ...815..116F}
{Fujita}, Y., {Takizawa}, M., {Yamazaki}, R., {Akamatsu}, H., and {Ohno}, H.
  (2015).
\newblock {Turbulent Cosmic-Ray Reacceleration at Radio Relics and Halos in
  Clusters of Galaxies}.
\newblock {\em \apj\/}, {\bf 815}, 116.

\bibitem[{Halverson} {\em et~al.}(2009){Halverson}, {Lanting}, {Ade}, {Basu},
  {Bender}, {Benson}, {Bertoldi}, {Cho}, {Chon}, {Clarke}, {Dobbs}, {Ferrusca},
  {G{\"u}sten}, {Holzapfel}, {Kov{\'a}cs}, {Kennedy}, {Kermish}, {Kneissl},
  {Lee}, {Lueker}, {Mehl}, {Menten}, {Muders}, {Nord}, {Pacaud}, {Plagge},
  {Reichardt}, {Richards}, {Schaaf}, {Schilke}, {Schuller}, {Schwan},
  {Spieler}, {Tucker}, {Weiss}, and {Zahn}]{2009ApJ...701...42H}
{Halverson}, N.~W., {Lanting}, T., {Ade}, P.~A.~R., {Basu}, K., {Bender},
  A.~N., {Benson}, B.~A., {Bertoldi}, F., {Cho}, H., {Chon}, G., {Clarke}, J.,
  {Dobbs}, M., {Ferrusca}, D., {G{\"u}sten}, R., {Holzapfel}, W.~L.,
  {Kov{\'a}cs}, A., {Kennedy}, J., {Kermish}, Z., {Kneissl}, R., {Lee}, A.~T.,
  {Lueker}, M., {Mehl}, J., {Menten}, K.~M., {Muders}, D., {Nord}, M.,
  {Pacaud}, F., {Plagge}, T., {Reichardt}, C., {Richards}, P.~L., {Schaaf}, R.,
  {Schilke}, P., {Schuller}, F., {Schwan}, D., {Spieler}, H., {Tucker}, C.,
  {Weiss}, A., and {Zahn}, O. (2009).
\newblock {Sunyaev-Zel'Dovich Effect Observations of the Bullet Cluster (1E
  0657-56) with APEX-SZ}.
\newblock {\em \apj\/}, {\bf 701}, 42--51.

\bibitem[{Kang}(2015){Kang}]{2015JKAS...48....9K}
{Kang}, H. (2015).
\newblock {Nonthermal Radiation From Relativistic Electrons Accelerated at
  Spherically Expanding Shocks}.
\newblock {\em Journal of Korean Astronomical Society\/}, {\bf 48}, 9--20.

\bibitem[{Kang}(2016){Kang}]{2016arXiv160307444K}
{Kang}, H. (2016).
\newblock {Re-acceleration model for the ''Toothbrush'' Radio Relic}.
\newblock {\em ArXiv e-prints\/}.

\bibitem[Kang and Ryu(2015)Kang and Ryu]{0004-637X-809-2-186}
Kang, H. and Ryu, D. (2015).
\newblock Curved radio spectra of weak cluster shocks.
\newblock {\em The Astrophysical Journal\/}, {\bf 809}(2), 186.

\bibitem[Kang and Ryu(2016)Kang and Ryu]{0004-637X-823-1-13}
Kang, H. and Ryu, D. (2016).
\newblock Re-acceleration model for radio relics with spectral curvature.
\newblock {\em The Astrophysical Journal\/}, {\bf 823}(1), 13.

\bibitem[{Liang} {\em et~al.}(2000){Liang}, {Hunstead}, {Birkinshaw}, and
  {Andreani}]{2000ApJ...544..686L}
{Liang}, H., {Hunstead}, R.~W., {Birkinshaw}, M., and {Andreani}, P. (2000).
\newblock {A Powerful Radio Halo in the Hottest Known Cluster of Galaxies 1E
  0657-56}.
\newblock {\em \apj\/}, {\bf 544}, 686--701.

\bibitem[{Malu} and {Subrahmanyan}(2011){Malu} and
  {Subrahmanyan}]{2011JApA...32..541M}
{Malu}, S.~S. and {Subrahmanyan}, R. (2011).
\newblock {18 GHz SZ Measurements of the Bullet Cluster}.
\newblock {\em Journal of Astrophysics and Astronomy\/}, {\bf 32}, 541--544.

\bibitem[{Malu} {\em et~al.}(2010){Malu}, {Subrahmanyan}, {Wieringa}, and
  {Narasimha}]{2010arXiv1005.1394M}
{Malu}, S.~S., {Subrahmanyan}, R., {Wieringa}, M., and {Narasimha}, D. (2010).
\newblock {Compact Sunyaev-Zeldovich `hole' in the Bullet Cluster}.
\newblock {\em ArXiv e-prints\/}.

\bibitem[{Owers} {\em et~al.}(2009){Owers}, {Nulsen}, {Couch}, and
  {Markevitch}]{2009ApJ...704.1349O}
{Owers}, M.~S., {Nulsen}, P.~E.~J., {Couch}, W.~J., and {Markevitch}, M.
  (2009).
\newblock {A High Fidelity Sample of Cold Front Clusters from the Chandra
  Archive}.
\newblock {\em \apj\/}, {\bf 704}, 1349--1370.

\bibitem[{Plagge} {\em et~al.}(2010){Plagge}, {Benson}, {Ade}, {Aird}, {Bleem},
  {Carlstrom}, {Chang}, {Cho}, {Crawford}, {Crites}, {de Haan}, {Dobbs},
  {George}, {Hall}, {Halverson}, {Holder}, {Holzapfel}, {Hrubes}, {Joy},
  {Keisler}, {Knox}, {Lee}, {Leitch}, {Lueker}, {Marrone}, {McMahon}, {Mehl},
  {Meyer}, {Mohr}, {Montroy}, {Padin}, {Pryke}, {Reichardt}, {Ruhl},
  {Schaffer}, {Shaw}, {Shirokoff}, {Spieler}, {Stalder}, {Staniszewski},
  {Stark}, {Vanderlinde}, {Vieira}, {Williamson}, and
  {Zahn}]{2010ApJ...716.1118P}
{Plagge}, T., {Benson}, B.~A., {Ade}, P.~A.~R., {Aird}, K.~A., {Bleem}, L.~E.,
  {Carlstrom}, J.~E., {Chang}, C.~L., {Cho}, H.-M., {Crawford}, T.~M.,
  {Crites}, A.~T., {de Haan}, T., {Dobbs}, M.~A., {George}, E.~M., {Hall},
  N.~R., {Halverson}, N.~W., {Holder}, G.~P., {Holzapfel}, W.~L., {Hrubes},
  J.~D., {Joy}, M., {Keisler}, R., {Knox}, L., {Lee}, A.~T., {Leitch}, E.~M.,
  {Lueker}, M., {Marrone}, D., {McMahon}, J.~J., {Mehl}, J., {Meyer}, S.~S.,
  {Mohr}, J.~J., {Montroy}, T.~E., {Padin}, S., {Pryke}, C., {Reichardt},
  C.~L., {Ruhl}, J.~E., {Schaffer}, K.~K., {Shaw}, L., {Shirokoff}, E.,
  {Spieler}, H.~G., {Stalder}, B., {Staniszewski}, Z., {Stark}, A.~A.,
  {Vanderlinde}, K., {Vieira}, J.~D., {Williamson}, R., and {Zahn}, O. (2010).
\newblock {Sunyaev-Zel'dovich Cluster Profiles Measured with the South Pole
  Telescope}.
\newblock {\em \apj\/}, {\bf 716}, 1118--1135.

\bibitem[Reynolds(1994)Reynolds]{reynolds1994}
Reynolds, J.~E. (1994).
\newblock A revised flux scale for the at compact array.
\newblock Technical Report 39.3/040, ATNF Technical Document Series.

\bibitem[{Sarazin}(1988){Sarazin}]{1988xrec.book.....S}
{Sarazin}, C.~L. (1988).
\newblock {\em {X-ray emission from clusters of galaxies}\/}.

\bibitem[Sault(2003)Sault]{atca_sault2003}
Sault, R. (2003).
\newblock {ATCA} flux density scale at 12mm.
\newblock Technical Report 39.3/124, ATNF Technical Document Series.

\bibitem[{Sault} {\em et~al.}(1996){Sault}, {Staveley-Smith}, and
  {Brouw}]{1996A&AS..120..375S}
{Sault}, R.~J., {Staveley-Smith}, L., and {Brouw}, W.~N. (1996).
\newblock {An approach to interferometric mosaicing.}
\newblock {\em \aaps\/}, {\bf 120}, 375--384.

\bibitem[{Shimwell} {\em et~al.}(2014){Shimwell}, {Brown}, {Feain}, {Feretti},
  {Gaensler}, and {Lage}]{2014MNRAS.440.2901S}
{Shimwell}, T.~W., {Brown}, S., {Feain}, I.~J., {Feretti}, L., {Gaensler},
  B.~M., and {Lage}, C. (2014).
\newblock {Deep radio observations of the radio halo of the bullet cluster 1E
  0657-55.8}.

\bibitem[{Shimwell} {\em et~al.}(2015){Shimwell}, {Markevitch}, {Brown},
  {Feretti}, {Gaensler}, {Johnston-Hollitt}, {Lage}, and
  {Srinivasan}]{2015MNRAS.449.1486S}
{Shimwell}, T.~W., {Markevitch}, M., {Brown}, S., {Feretti}, L., {Gaensler},
  B.~M., {Johnston-Hollitt}, M., {Lage}, C., and {Srinivasan}, R. (2015).
\newblock {Another shock for the Bullet cluster, and the source of seed
  electrons for radio relics}.

\bibitem[{Siemieniec-Ozieblo}(2004){Siemieniec-Ozieblo}]{2004AcPPB..35.2131S}
{Siemieniec-Ozieblo}, G. (2004).
\newblock {A Study of Diffusive Shock Acceleration as a Process Explaining
  Observations of 1 E0657-56 Galaxy Cluster}.
\newblock {\em Acta Physica Polonica B\/}, {\bf 35}, 2131.

\bibitem[{Stroe} {\em et~al.}(2014){Stroe}, {Rumsey}, {Harwood}, {van Weeren},
  {R{\"o}ttgering}, {Saunders}, {Sobral}, {Perrott}, and
  {Schammel}]{2014MNRAS.441L..41S}
{Stroe}, A., {Rumsey}, C., {Harwood}, J.~J., {van Weeren}, R.~J.,
  {R{\"o}ttgering}, H.~J.~A., {Saunders}, R.~D.~E., {Sobral}, D., {Perrott},
  Y.~C., and {Schammel}, M.~P. (2014).
\newblock {The highest frequency detection of a radio relic: 16 GHz AMI
  observations of the `Sausage' cluster}.
\newblock {\em \mnras\/}, {\bf 441}, L41--L45.

\bibitem[{Stroe} {\em et~al.}(2016){Stroe}, {Shimwell}, {Rumsey}, {van Weeren},
  {Kierdorf}, {Donnert}, {Jones}, {R{\"o}ttgering}, {Hoeft},
  {Rodr{\'{\i}}guez-Gonz{\'a}lvez}, {Harwood}, and
  {Saunders}]{2016MNRAS.455.2402S}
{Stroe}, A., {Shimwell}, T., {Rumsey}, C., {van Weeren}, R., {Kierdorf}, M.,
  {Donnert}, J., {Jones}, T.~W., {R{\"o}ttgering}, H.~J.~A., {Hoeft}, M.,
  {Rodr{\'{\i}}guez-Gonz{\'a}lvez}, C., {Harwood}, J.~J., and {Saunders},
  R.~D.~E. (2016).
\newblock {The widest frequency radio relic spectra: observations from 150 MHz
  to 30 GHz}.
\newblock {\em \mnras\/}, {\bf 455}, 2402--2416.

\bibitem[{Subrahmanyan} {\em et~al.}(2000){Subrahmanyan}, {Kesteven}, {Ekers},
  {Sinclair}, and {Silk}]{2000MNRAS.315..808S}
{Subrahmanyan}, R., {Kesteven}, M.~J., {Ekers}, R.~D., {Sinclair}, M., and
  {Silk}, J. (2000).
\newblock {An Australia Telescope survey for CMB anisotropies}.
\newblock {\em \mnras\/}, {\bf 315}, 808--822.

\bibitem[{van Weeren} {\em et~al.}(2010){van Weeren}, {R{\"o}ttgering},
  {Br{\"u}ggen}, and {Hoeft}]{2010Sci...330..347V}
{van Weeren}, R.~J., {R{\"o}ttgering}, H.~J.~A., {Br{\"u}ggen}, M., and
  {Hoeft}, M. (2010).
\newblock {Particle Acceleration on Megaparsec Scales in a Merging Galaxy
  Cluster}.
\newblock {\em Science\/}, {\bf 330}, 347.

\bibitem[{van Weeren} {\em et~al.}(2012){van Weeren}, {R{\"o}ttgering},
  {Intema}, {Rudnick}, {Br{\"u}ggen}, {Hoeft}, and {Oonk}]{2012A&A...546A.124V}
{van Weeren}, R.~J., {R{\"o}ttgering}, H.~J.~A., {Intema}, H.~T., {Rudnick},
  L., {Br{\"u}ggen}, M., {Hoeft}, M., and {Oonk}, J.~B.~R. (2012).
\newblock {The ''toothbrush-relic'': evidence for a coherent linear 2-Mpc scale
  shock wave in a massive merging galaxy cluster?}
\newblock {\em \aap\/}, {\bf 546}, A124.

\bibitem[{Wilson} {\em et~al.}(2011){Wilson}, {Ferris}, {Axtens}, {Brown},
  {Davis}, {Hampson}, {Leach}, {Roberts}, {Saunders}, {Koribalski}, {Caswell},
  {Lenc}, {Stevens}, {Voronkov}, {Wieringa}, {Brooks}, {Edwards}, {Ekers},
  {Emonts}, {Hindson}, {Johnston}, {Maddison}, {Mahony}, {Malu}, {Massardi},
  {Mao}, {McConnell}, {Norris}, {Schnitzeler}, {Subrahmanyan}, {Urquhart},
  {Thompson}, and {Wark}]{2011MNRAS.416..832W}
{Wilson}, W.~E., {Ferris}, R.~H., {Axtens}, P., {Brown}, A., {Davis}, E.,
  {Hampson}, G., {Leach}, M., {Roberts}, P., {Saunders}, S., {Koribalski},
  B.~S., {Caswell}, J.~L., {Lenc}, E., {Stevens}, J., {Voronkov}, M.~A.,
  {Wieringa}, M.~H., {Brooks}, K., {Edwards}, P.~G., {Ekers}, R.~D., {Emonts},
  B., {Hindson}, L., {Johnston}, S., {Maddison}, S.~T., {Mahony}, E.~K.,
  {Malu}, S.~S., {Massardi}, M., {Mao}, M.~Y., {McConnell}, D., {Norris},
  R.~P., {Schnitzeler}, D., {Subrahmanyan}, R., {Urquhart}, J.~S., {Thompson},
  M.~A., and {Wark}, R.~M. (2011).
\newblock {The Australia Telescope Compact Array Broad-band Backend:
  description and first results}.
\newblock {\em \mnras\/}, {\bf 416}, 832--856.

\end{thebibliography}

\end{document}